# Towards a Microservice-based Middleware for a Multi-hazard Early Warning System

Adeyinka Akanbi[1][0000-0002-8796-0674]

[1] Centre for Sustainable Smart Cities (CSSC), Central University of Technology, Free State, 9300, South Africa
aakanbi@cut.ac.za

**Abstract.** Environmental hazards—like water and air pollution, extreme weather, or chemical exposures—can affect human health in a number of ways, and it is a persistent apprehension in communities surrounded by mining operations. The application of modern technologies in the environmental monitoring of these Human-made hazards is critical, because while not immediately health-threatening may turn out detrimental with unwanted negative effects. Enabling technologies needed to realize this concept is multifaceted and most especially involves deploying interconnected Internet of Things (IoT) sensors, existing legacy systems, enterprise networks, multi-layered software architecture (middleware), and event-processing engines, amongst others. Currently, the integration of several early warning systems has inherent challenges, mostly due to the heterogeneity of components. This paper proposes transversal microservice-based middleware aiming at increasing data integration, interoperability, scalability, high availability, and reusability of adopted systems using a containers orchestration framework for a multi-hazard early warning system. Devised within the scope of the ICMHEWS project, the proposed platform aims at improving known challenges.

## 1 Introduction

Natural hazards can be defined as "*a serious disruption of the functioning of a community or a society causing widespread human, material, economic or environmental losses which exceed the ability of the affected community or society to cope using its own resources*" [1]. The preparedness towards natural hazards is a key factor in the reduction of their impact on society. Natural hazards/disasters are mostly from compromised hydro-meteorological origins resulting in pollution and chemical exposure to naturally occurring ones from extremes of temperature, wind and rainfall. An important part of a holistic approach to disaster risk reduction (DRR) management of natural hazards or disasters is the set-up of early warning systems, with several international initiatives towards the development and promotion of early warning systems for all natural hazards [3],[4],[5].

Early warning systems (EWS) can be defined as information systems with the ability to detect and provide warnings in the form of timely and effective information through identified institutions that allow individuals exposed to a hazard to take action to avoid or reduce their risk and prepare for effective response [1]. Several studies have illustrated the effectiveness of an early warning system (e.g., [8],[9],[12],[13],).

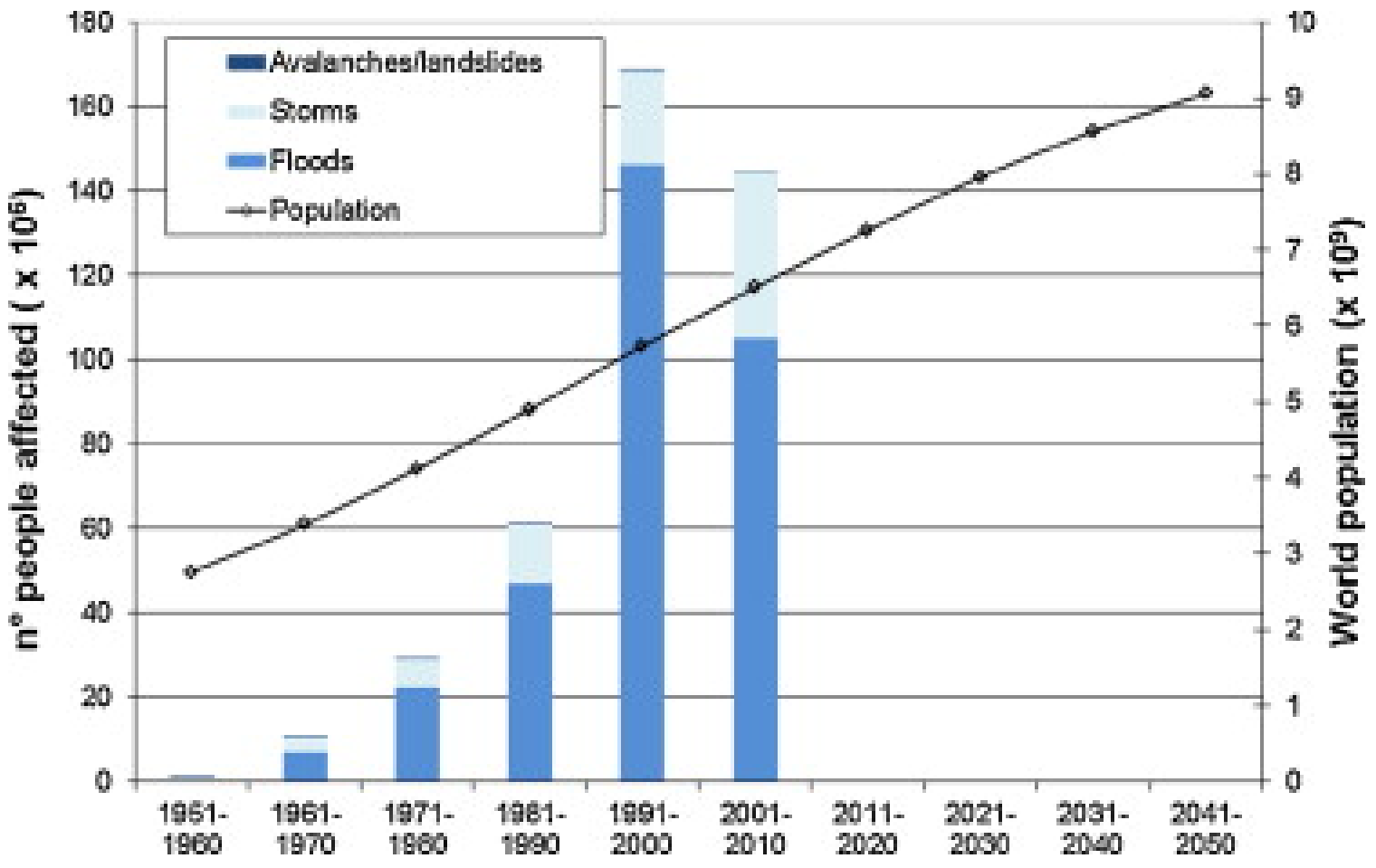

**Fig. 1.** Mean yearly number of people affected per decade (stacked bars) and comparison with global population for global water-related disasters in the world in the last 60 years, including trend for the future 40 years [6]

In previous studies, EWS(s) are primarily constructed to target a particular natural hazard – this approach is widely used, especially those concerning natural hazards [7],[10],[14],[15],[16]. However, the world's climate is changing [17-19], and natural hazards are now intertwined with environmental phenomena leading directly/indirectly to natural hazards occurring concurrently with cascading effects; the failure to integrate EWS could affect their effectiveness and reach. Thus, it is important to devise the integration of new and existing EWS for an integrated climate multi-hazard early warning system (ICMHEWS[1]). The specifications and requirements of existing EWS are extremely different depending on the application area, leading to ad-hoc implementations of monolithic software applications and cumbersome legacy systems. The integration of several related EWS has inherent challenges, which are mostly data incompatibility and system interoperability due to heterogeneity, reliability, availability, transparency and abstractions to applications [2],[20-21],[35-36], inhibiting the possibility of harnessing an integrated MHEWS.

[1] https://urida.co.za/icmhews

In recent years, the field of cloud computing has shown rapid growth, and a variety of virtualization technologies have emerged, such as microservices, with immense application characteristics for monolithic heterogeneous systems or applications. Monolithic applications components are tightly coupled, having been developed, deployed and managed as one entity. This results in increased rigidity and complexity of the system. On the other hand, microservices are loosely coupled, independently deployed, cloud-native small services [22]. The cloud-native microservices exploit containerization orchestration framework and container management systems such as Google Kubernetes to deploy software components or applications separately, without compromising the application life cycle [23],[26]. Containers encapsulate a microservice environment, abstracting the hardware and software infrastructure and provide application portability across platforms as a resource-isolated process. This provides the ability to break down monolithic applications into software components, and run them as a node on a variety of Infrastructure as a Service (IaaS) or Platform as a Service (PaaS). Therefore, containerization enables a paradigm shift from machine-oriented to application-oriented orchestration, resulting in easier and faster deployment, improved scalability, increased utilization of computing resources, data integration and system interoperability. To automatically manage applications with containers, several orchestration frameworks are developed, such as Kubernetes [26], Docker SwarmKit [25] and Apache Mesos [11].

In this paper, we develop a formal model towards the decomposition of monolithic EWS components as containerized microservices managed by Kubernetes. This allows the deployment of EWS software components towards an integrated MHEWS under several configurations to be explored at the modelling level before deployment to production. Thus, the analysis of the performance, suitability, and usability of Kubernetes in a decoupled monolithic EWS is an interesting and relatively new research area. We aim to facilitate the expediency of the Kubernetes container orchestration tool in MHEWS and highlight the limitation therein.

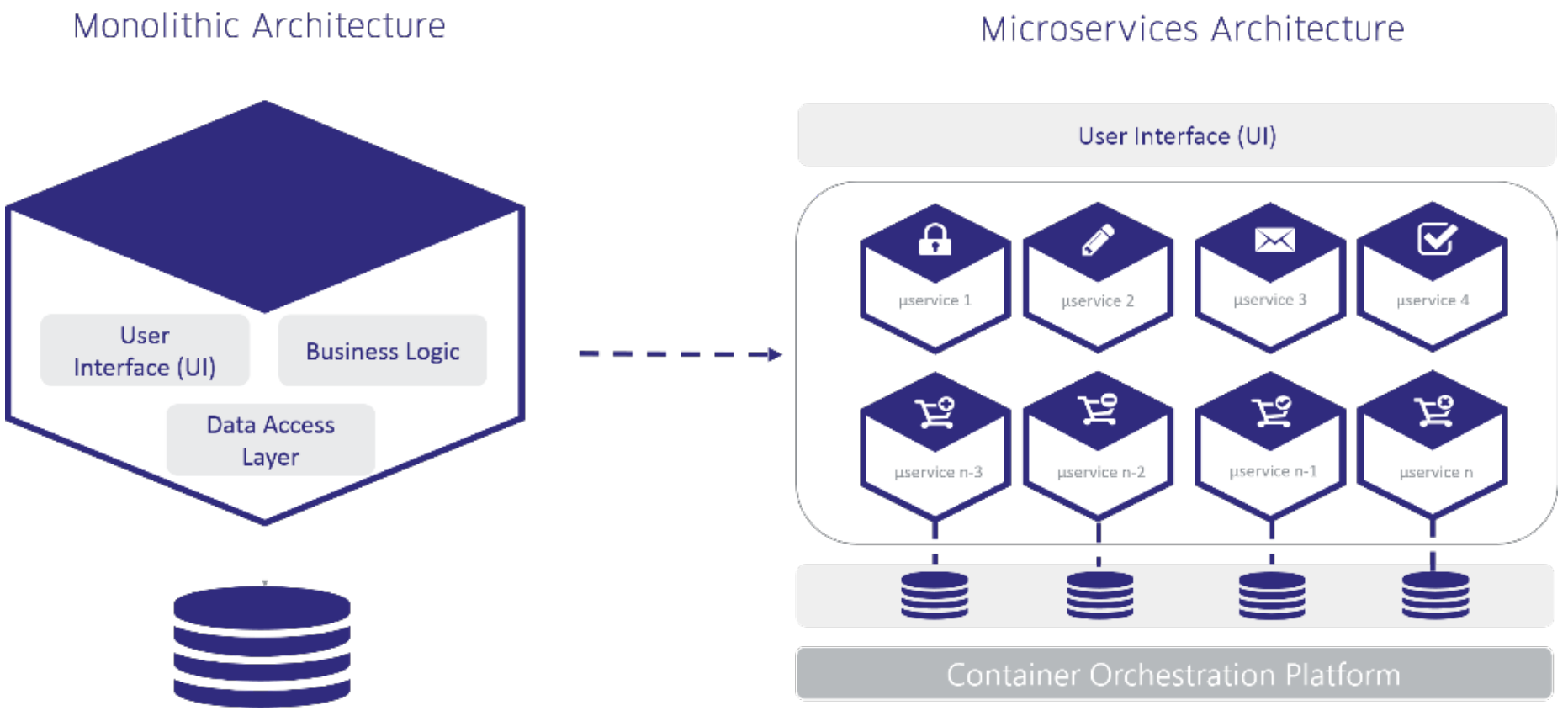

**Fig. 2.** Monolithic vs Microservice Architecture

More specifically, the contributions of this research study can be summarized as follows: (i) describing the model for the distribution of containerized EWS software applications; (ii) proposing a model for the application of Kubernetes container scheduling techniques for the deployment of reliable and scalable MHEW distributed systems; (iii) validation in EWS is conducted, more especially, in drought forecasting domain, (iv) discussing the limitation of current Kubernetes container orchestration design for EWS.

The rest of the paper is organized as follows. In the next Section, the background is discussed. Section 3 presents the proposed experimental framework architecture, the implementation in a test environment and performance results. Finally, conclusions are presented in Section 4.

## 2 Background

In this section, we present the relevant background of our work; we start by presenting an overview of microservices and container management. Then we present the containerized orchestration framework for the study.

### 2.1 Microservices and Containers Management

In a monolithic software application, all components and services are highly coupled, preventing scalability and reusability of these systems or even integration with new or existing ones. However, to overcome these challenges, a microservices-based architecture is used. The application principle of microservices is all about modularization and decoupling capabilities into components that are easily adapted to distributed hardware. This emanates from service-oriented architectures (SOA) [24]. In a nutshell, microservice-based architecture is the evolution of classical SOA [22, 33]. The adaptability of the SOA approach to a transversal microservice-based middleware is to ensure seamless implementations of the various software component such as APIs, extensions, heterogeneous technologies or clusters in the monolithic software application or systems.

Microservices are independent components conceptually deployed in isolation and equipped with dedicated resources for utilization. The components of a microservice architecture are microservices, with different behaviour derives from the composition and coordination of its decoupled software components. Microservices manage growing complexity by functionally decomposing large systems into a set of independent services [22]. This takes modularity to the next level by making services completely independent in development and deployment, through emphasis on loose coupling and high cohesion. This approach delivers all sorts of benefits in terms of maintainability, scalability, integration and interoperabilty. Containers encapsulate the execution environment providing the ability to develop, deploy and scale applications as multiple instances or a set of services without dependencies [22].

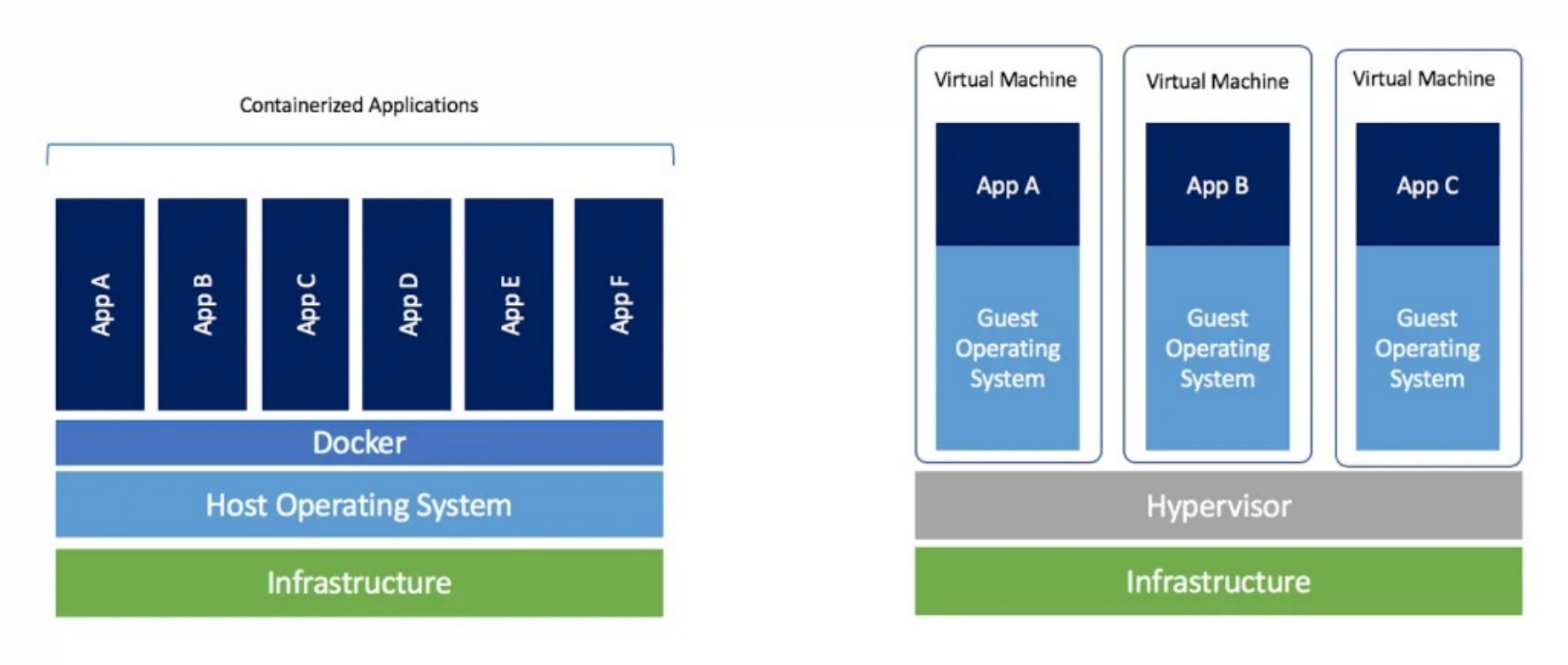

**Fig. 3.** A comparison of model of containerized application and VMs.

In literature, there are several container orchestration tools developed by different companies or open-source communities, typical examples such as Google Kubernetes [26], Apache Mesos [11], OpenShift [27], Nomad [28], Docker Compose [29], Cloudify [30], etc. Google Kubernetes is an open-source container orchestration tool for managing containerized applications across multiple hosts [23]. It provides automatic deployment, scaling and management of container-based applications or software components. Figure 3 depicts a logical representation of Kubernetes instances. The architectures follow a master-slave model or Pods concept, where a master node manages the worker nodes (slaves) – set up as a cluster consisting of Kubernetes-master and a set of Kubernetes-workers. As consequence, these nodes can be executed on-premises, in public cloud or hybrid infrastructure. The communication between microservices is possible only through interfaces using APIs.

There are four master processes in a Kubernetes-master node, namely: the *API server*, *scheduler*, *controller manager* and *etcd*. The Kubernetes-worker node has three processes, the *container runtime*, *Kubelet* and the *KubeProxy*. The container runtime needs to be installed on every node. The smallest unit of a Kubernetes cluster is a pod, which is an abstraction over the container runtime; usually, one application is dedicated to running in pod. The communication between pods is possible through virtual networks, with each pod having its own internal IP address. Within one pod, containers can reference each other directly. The access to the executed applications is through external service in the form of the node IP address and the service's port number e.g., `http://124.95.101.2:8080`. The external request goes to Ingress, which passes the request to the services residing in the container node.

Services are an integral part and another component of Kubernetes; services comprise a static or permanent IP address attached to each pod and act like a load balancer between pods. Each app in a pod has its own respective service with a disjointed life cycle. The two sub-types of services are the internal service and the external service. The internal service received request from ingress to access running containerized applications via respective endpoints. The orchestration of requests is possible with the help of *ConfigMap*, which contains the external configuration of applications, they are connected to the pods for integrated applications. Configuration of secured external applications makes use of *Secret*, which is similar to the *ConfigMap* to store access credentials for secured infrastructure. Volumes are another important feature that allows saving of persistent data required by applications running in the pods. These data are available through external storage attached physically to the infrastructure in an on-prem environment or remotely to the cloud infrastructure.

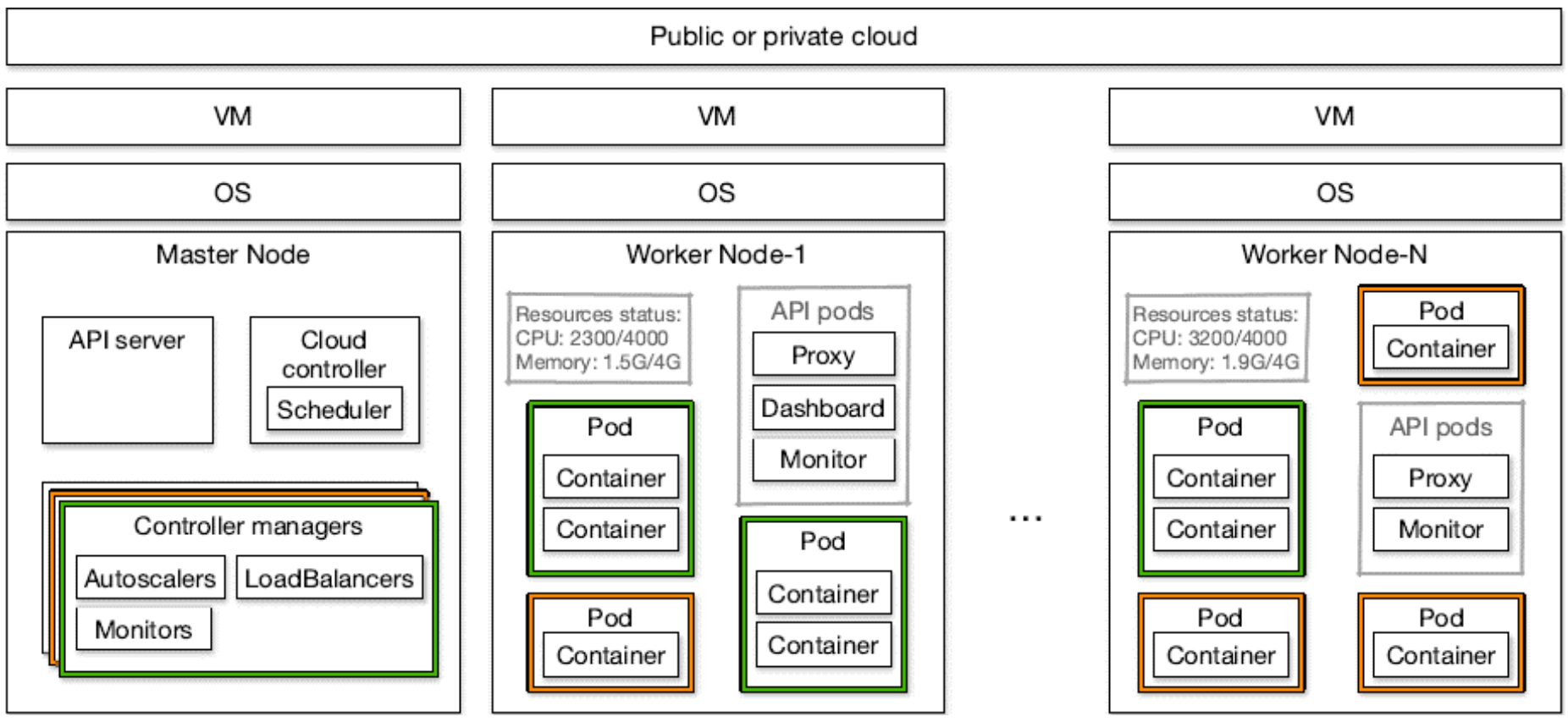

**Fig. 4.** A logical representation of Kubernetes components in a generic infrastructure [37].

## 2.2 Container Orchestration

Containerization expedites the feasibility of running applications that are containerized over multiple hosts in different service models [31]. Kubernetes has grown into container orchestration standards by simplifying the deployment and management of a containerized application. The Kubernetes-master provides the API server – a cluster gateway for scheduling various deployments and managing the overall cluster. This is achieved through RESTFul interface, which allows control point for managing the entire Kubernetes cluster. The interactions with the clusters or configuration of the Kubernetes-worker nodes are through *Kubectl* – a built-in Kubernetes Command Line Interface (CLI). The *scheduler* receives validated requests from the API server to start pods in the cluster. The *container manager* detects the state changes and notifies the *scheduler* if a container has to be restarted. *Etcd* is a key value pair store of the cluster state, used for coordinating resources and sharing cluster configuration; it acts as the brain of the cluster. In the Kubernetes-workers node, The *Kubelet* is a process that interfaces with both the container and the node and is responsible for starting a pod within a container and assigning resources from the node to the container. *KubeProxy* forwards service requests intelligently to available replicas in the cluster.

A cluster orchestration platform should be able to have fully automated, self-managed and self-healing capabilities. It also provides the ability for scalability and integration of containerized applications, promoting interoperability and eliminating the isolation of applications and systems. Among various available orchestration platforms in this paper, we have used Kubernetes for monitoring and managing EWS software components or applications.

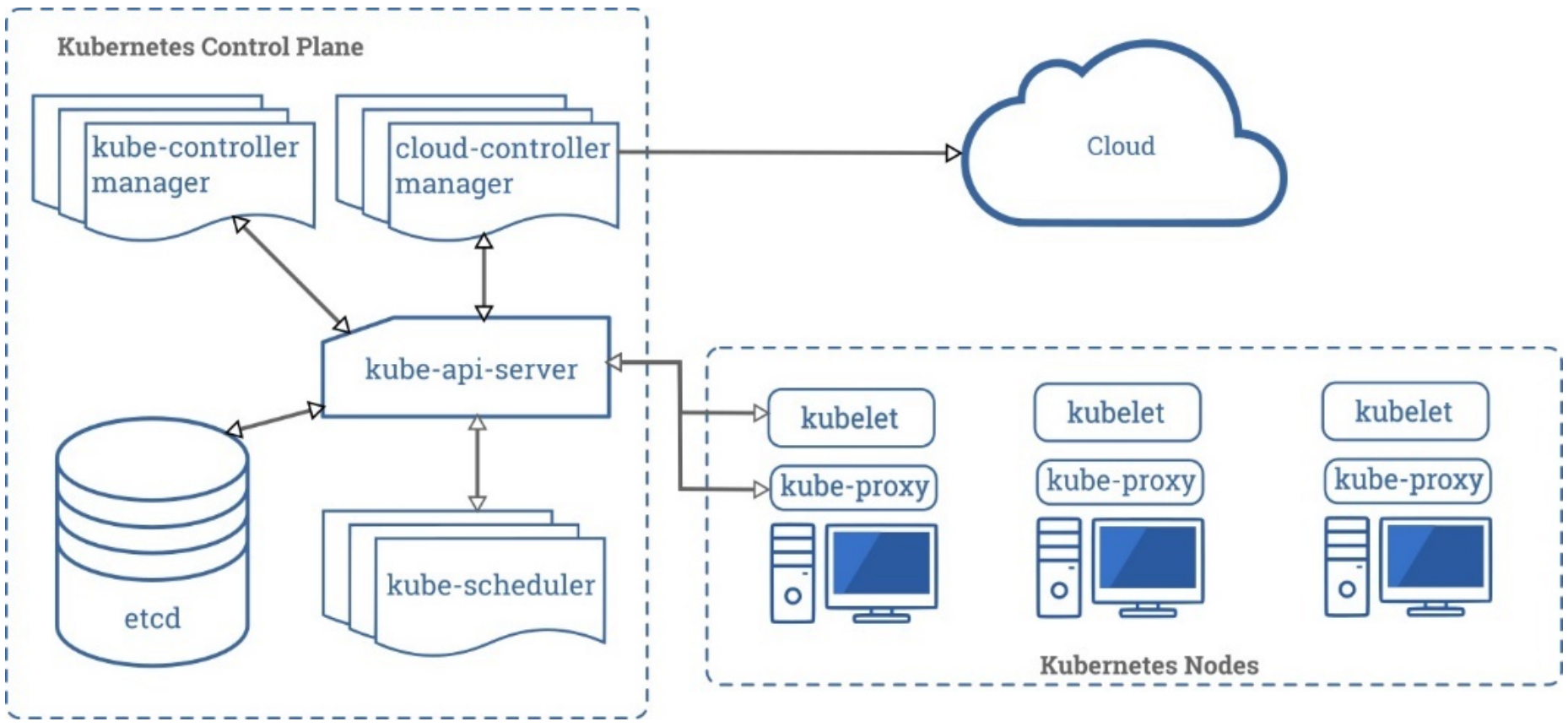

**Fig. 5.** Components of Kubernetes [38].

## 3 Proposed Experimental Framework

In this study, we presented an experimental framework towards the integration of several EWS for an integrated MHEWS using microservices. The objective is to address and eliminate the rigidity of monolithic EWS application for an ICMHEWS by implementing Kubernetes in a hybrid infrastructure. The infrastructure design consists of on-premises workstation and VMs in the cloud. The study adopts Microsoft Azure[2] cloud services to host the VMs with Azure Kubernetes Services, which are accessible via a public endpoint. Azure Kubernetes Services[3] (AKS) provides automated management and scalability of Kubernetes clusters for our container orchestration with the ability to deploy containerized Windows and Linux applications in the cloud. Kubernetes orchestrates clusters of VMs and schedules containers to run on those virtual machines based on available resources and the resource requirements of each container. The presented solution suggests the ability of Kubernetes to implement containerized EWS applications using it load-balancing capabilities to respond to requests.

[2] https://www.portal.azure.com
[3] https://azure.microsoft.com/en-us/products/kubernetes-service/

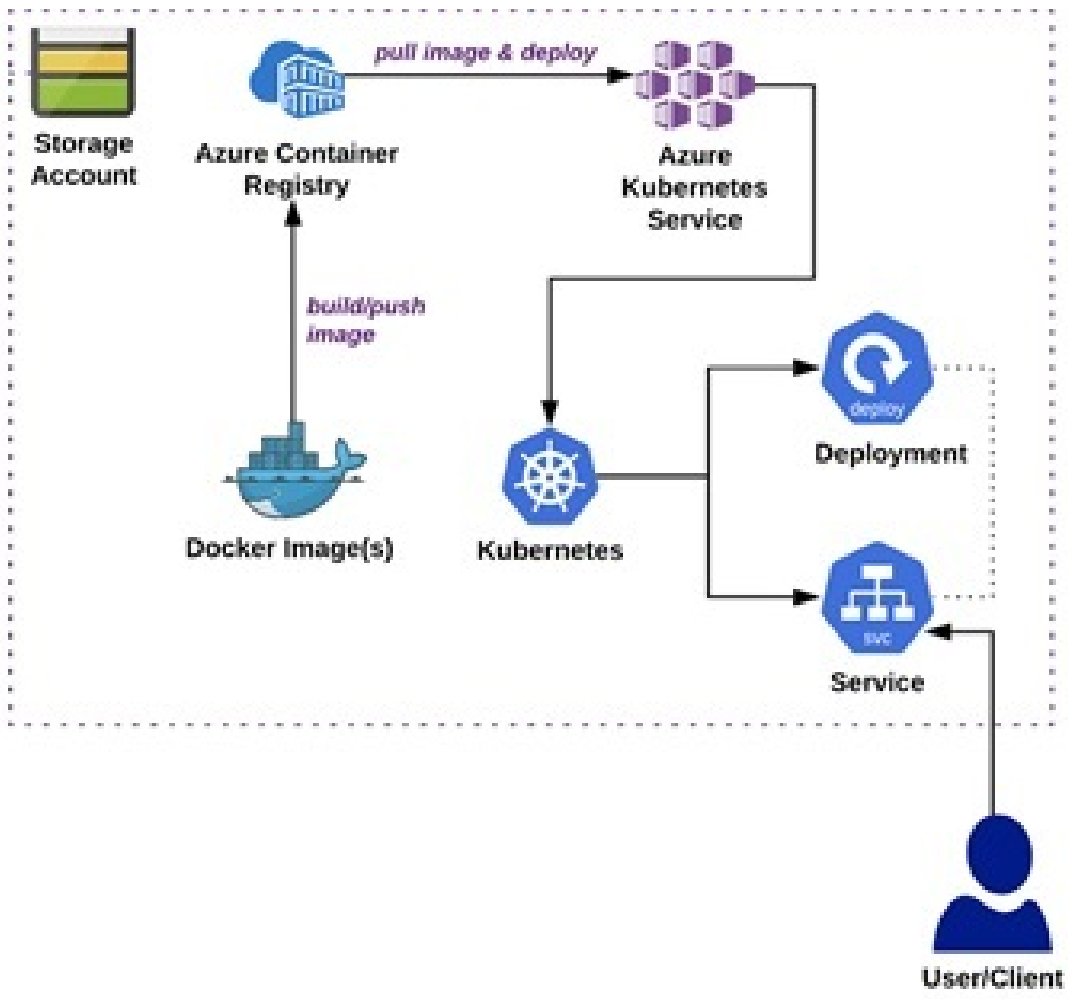

**Fig. 6.** Experimental framework of EWS application as microservices in AKS.

### 3.1 Cluster Setup

To replicate the execution of EWS application, we converted a software component of an EWS into a docker base image to be deployed as a containerized application using AKS. Our testbed is a VM in the Azure cloud, using a preset cluster configuration for the testing/dev environment. The VM configurations utilize a 4 vCPUs and 16GB memory with a primary node pool for size 10. All nodes run Kubernetes version 1.23.12. The cluster name is given as MHEWS_KB_Cluster with a default scaling setting at Autoscale for prompt scalability depending on the computing requirement of the executed containerized application. The complete environmental variables for the cluster are available on GitHub at [34]. After creating the docker images now to orchestrate these containers created using installed Kubernetes 1.23.12 in the cloud cluster. The master node is the principal node controlling the rest of the machines which run as container execution nodes. Kubernetes provides the tools that automate the distribution of applications across the cluster. Next, we configure Kubernetes to deploy the application for conducting the experiment and compare the performance result. Figure 7, 8 & 9 below depicts the pods in the cluster nodes, the up-running services that facilitate communication and provision of requested services through the endpoints and the external service endpoint to access the cluster through the BASH shell.

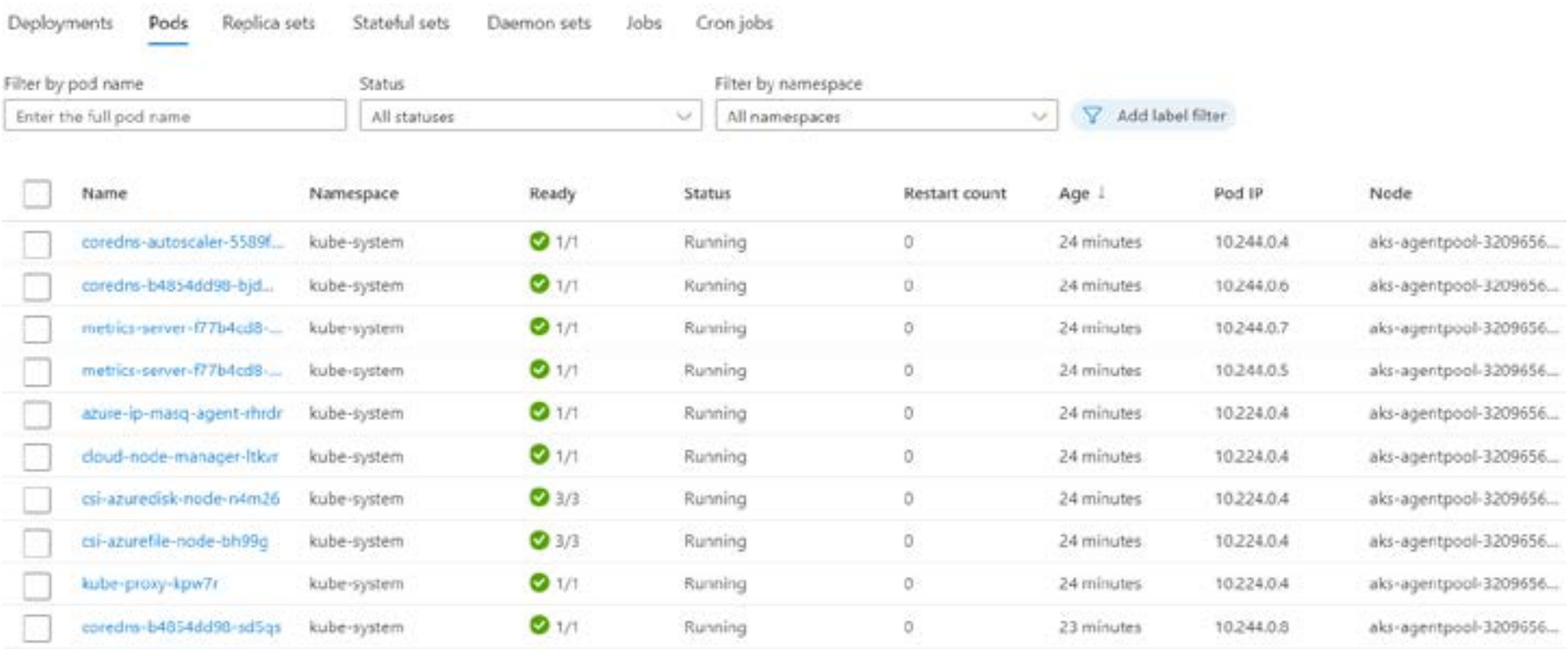

**Fig. 7.** Running nodes' pods in the deployed cluster

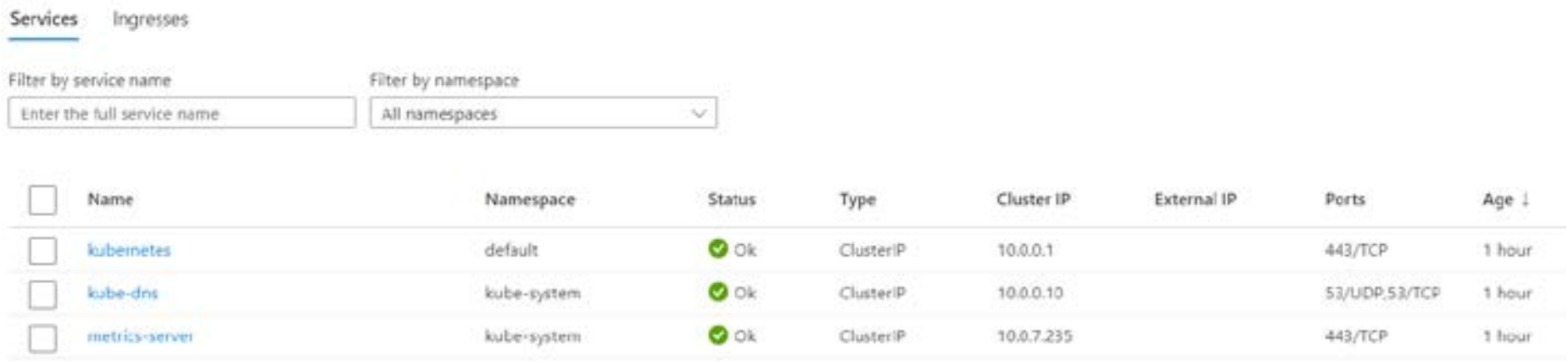

**Fig. 8.** Deployed Kubernetes services and ingresses

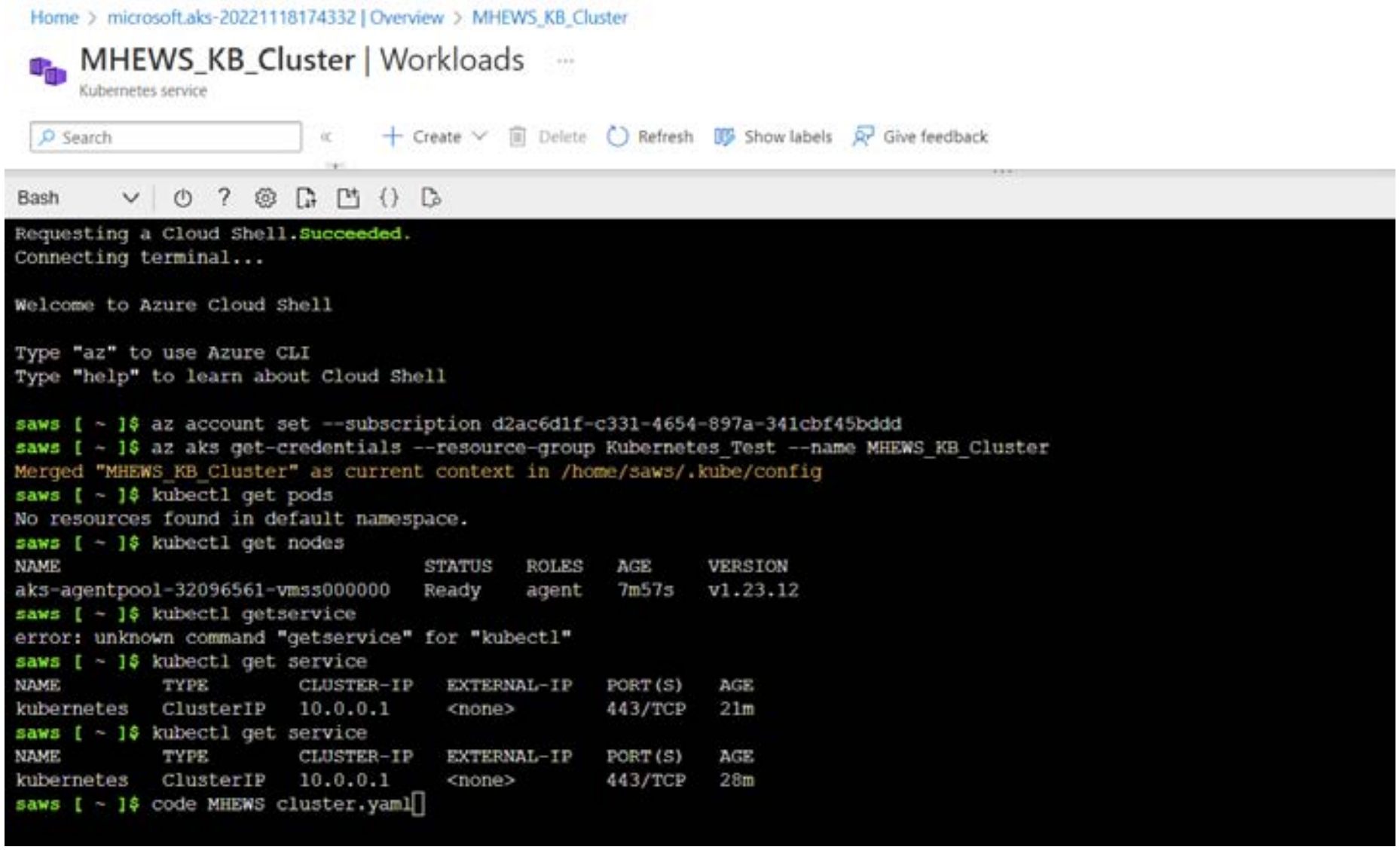

**Fig. 9.** Verification of the external service endpoint to access the cluster through the BASH shell.

A pod is defined by the YAML file that consists of the parameters of the container image for the EWS application. Code snippet 1 shows the example of variables for creating a microservice pod for Figure 6.

**Listing 1.** Creating Pod for MHEWS microservice.

```
apiVersion: apps/v1
kind; Deployment
metadata
  name: MHEWS cluster
spec:
  replicas:1
  selector:
    matchLabels:
      app: MHEWS Cluster
  template:
    metadata:
      labels:
        app: MHEWS Cluster
```

To enable the monitoring of the executed containerized application, the performance of the deployed microservice was monitored against the compute resources provided for the cluster. The following resource utilization metrics were considered:

- Throughput, which is a measure of how many units of information a system can process in a given amount of time is measured as a performance metrics of the cluster. It is measured in bits / second.
- Cluster performance is the CPU utilization when interacting with the cluster. It is measured in CPU core usage in milliseconds and percentages.
- Other metrics obtained are memory usage, network utilization in bytes and statuses for various node conditions.

### 3.2 Results

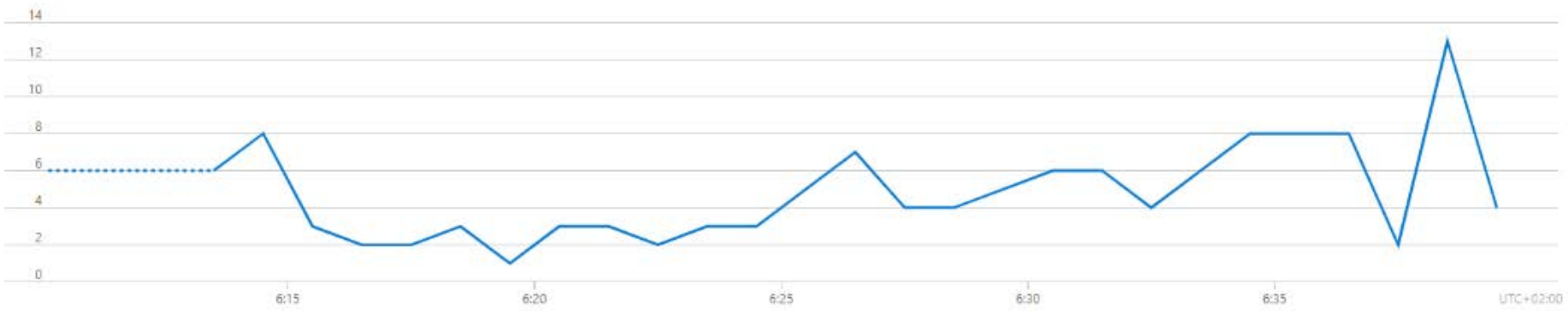

**Fig. 10.** Average Throughput for deployed microservice.

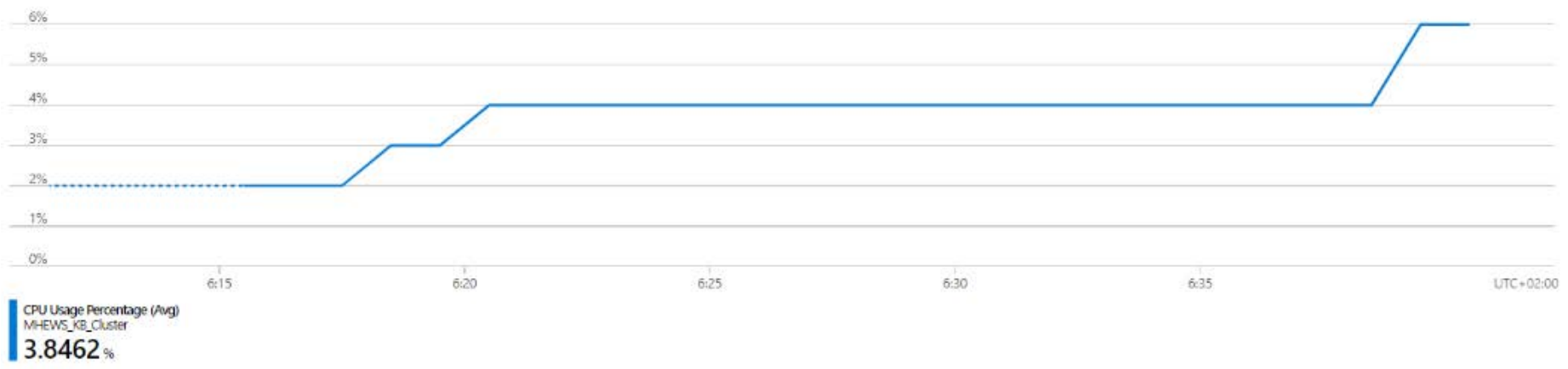

**Fig. 11.** CPU usage utilization.

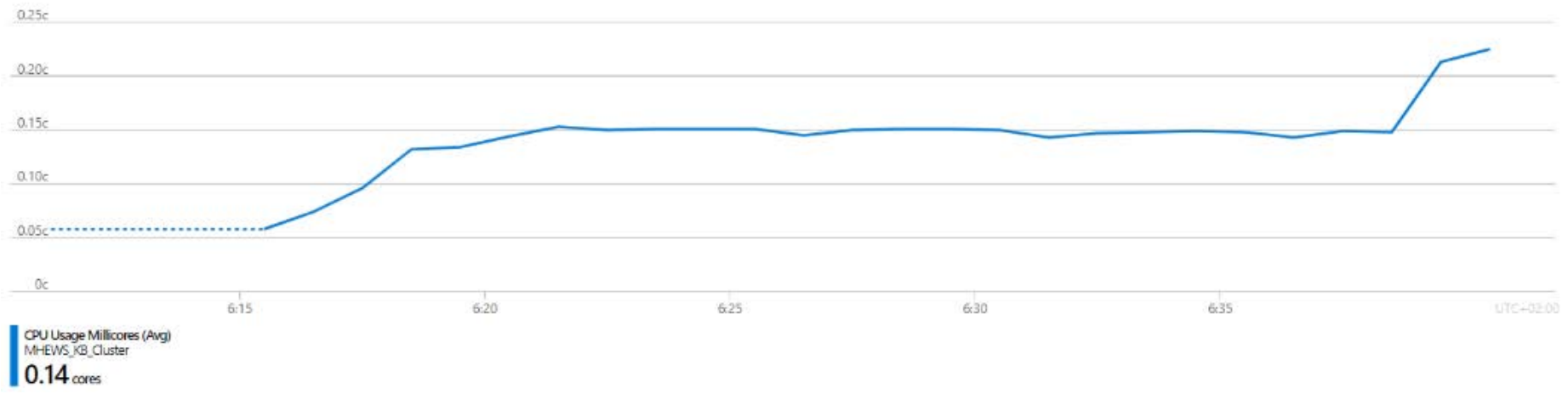

**Fig. 12.** Average CPU usage Milli-cores.

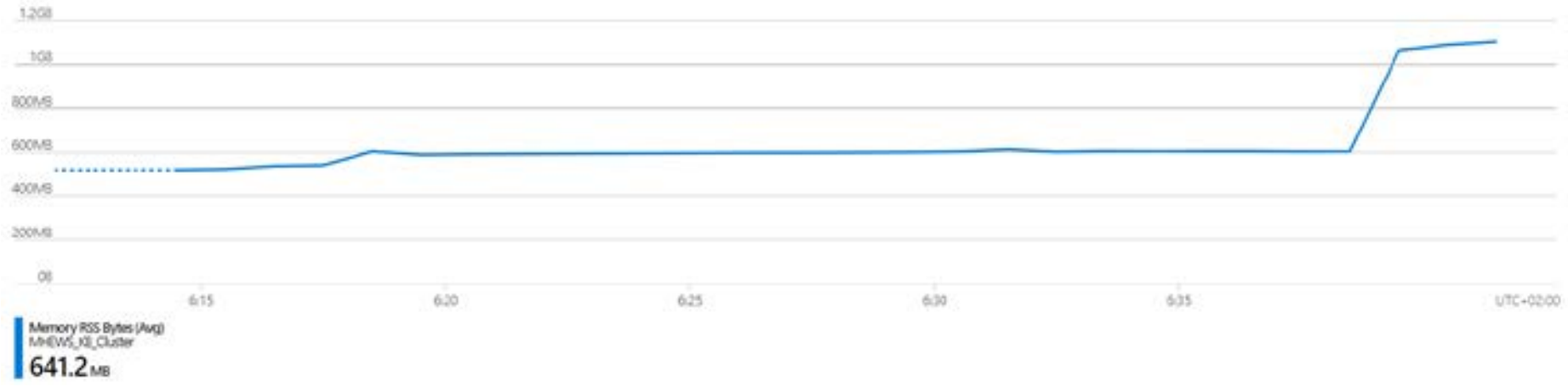

**Fig. 13.** Average cluster memory utilization.

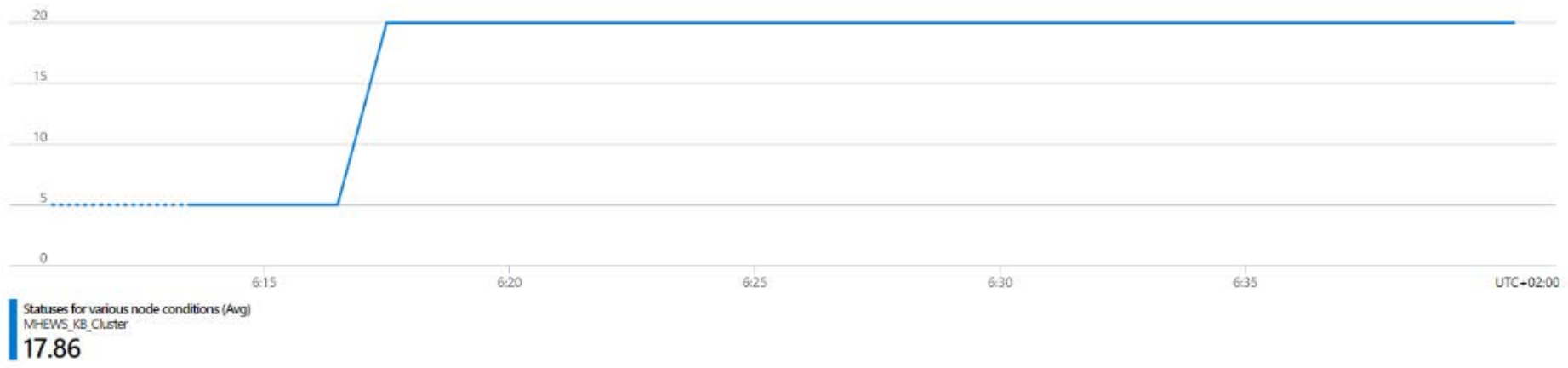

**Fig. 14.** Status of various node conditions

Figures 10, 11 and 12 show the effect of executing the containerized application as a microservice in the cluster. The figures reveal the throughput and CPU usage on average during execution, which are not overloaded or above the median threshold for the cluster configuration resources. This is important as it shows the ability of Kubernetes to run EWS applications as a containerized microservice. Figures 13 and 14 show a minimum memory utilization of the cluster – all within limits. From the results shows,

we can conclude that Kubernetes microservices is an excellent choice for decoupling monolithic EWS applications towards integration of several EWS for an integrated climate-driven multi-hazard early warning system.

The experimental results discussed above show that containers enabled with Kubernetes are capable of running the EWS application as a microservice, which will foster integration and interoperability towards a fully-fledged integrated climate multi-hazard early warning system. The experimental setup to validate the proposed framework proved to be advantageous.

# 4 Conclusion

In this study, we presented an experimental framework towards a middleware that integrates several EWS for an integrated climate-driven multi-hazard early warning system using Kubernetes microservices. The experiment shows proper execution of the deployed EWS docker image in the pod. The deployed pods were monitored based on the throughput and CPU utilizations to verify the ability of pods to run containerized applications and has performed optimally. Preliminary tests carried out on the platform are encouraging, but there are still much work to do in many aspects. This study is advantageous in light of a research study that predicts over 75% of global organizations are expected to run containerized applications in production by 2022-2023 [32]. The work reported in this paper is a subset of a bigger project aimed at increasing data integration, interoperability, scalability, high availability, and reusability of EWSs for an integrated climate-driven multi-hazard early warning system.